\documentclass[11pt,twoside]{article}

\usepackage{asp2004}
\usepackage{epsf}
\usepackage{psfig}
\usepackage{lscape}
\usepackage{graphicx}

\begin{document}
\title{The IRAC galaxy correlation functions from SWIRE}   
\author{Ian Waddington$^1$ and the SWIRE Team}   
\affil{{}$^1$Astronomy Centre, University of Sussex, Brighton, BN1 9QH, UK.}    

\begin{abstract} 
We present an analysis of large-scale structure from the Spitzer
Wide-area Infrared Extragalactic legacy survey, SWIRE.  The two-point
angular correlation functions were computed for galaxies detected in
the 3.6-micron IRAC band, on angular scales up to a degree.
Significant evolution in the clustering amplitude was detected, as the
median redshift of the samples increases from $z=0.2$ to 0.6.  The
galaxy clustering in the {\sc galics} semi-analytic models was
compared with the observed correlation functions and found to disagree
with the data at faint flux limits.
\end{abstract}



\section{Introduction}

The large-scale structure of the universe is a result of the
gravitational growth of dark matter density perturbations.  The
detailed structure is dependent on both the cosmological parameters,
which can be constrained by measurements of the cosmic microwave
background for example, and the evolution of the dark matter
distribution, as modelled with N-body simulations and semi-analytic
techniques.  It is known that the distribution of galaxies is biased
with respect to the dark matter, but the details of the bias is poorly
constrained.  Observations of the large-scale structure of galaxies,
and its evolution, can be used to understand this relationship between
galaxy formation and the mass field.

The Spitzer Wide-area Infrared Extragalactic survey
\citep[SWIRE;][]{lonsdale03,lonsdale04} is the largest of the Spitzer
Space Telescope's six legacy programmes.  When completed, we will have
observed a total area of 49~square degrees, split between six fields,
in all seven of Spitzer's imaging bands (3.6--160~\micron).  SWIRE is
being used to study the history of star formation and the assembly of
stellar mass in galaxies, the nature and impact of accretion in active
galactic nuclei, and the influence of environment on these processes.

The area and depth of SWIRE combine to produce a survey of significant
cosmological volume.  Out to the estimated median redshift of $z=1$,
the survey proper volume is 0.04~Gpc$^3$ -- larger than the Sloan
Digital Sky Survey (DR3 spectroscopic sample, median $z=0.1$;
\citeauthor{Abazajian05}~2005).  SWIRE is sensitive to both
star-forming and passively-evolving galaxies in the same volume,
extending out to redshifts of 2--3 and across comoving spatial scales
up to 100~Mpc.  In \citet{Oliver04} we presented the first detection
of galaxy clustering with Spitzer, measuring a two-point angular
correlation function at 3.6~\micron\ from our validation data.
Similarly, \citet{Fang04} presented their IRAC angular correlation
functions from the Spitzer First Look Survey.  Here we begin to extend
our work to the larger SWIRE survey.

\section{The two-point angular correlation functions}

The restframe emission of galaxies in the $H$- and $K$-bands is
dominated by older stellar populations and traces the stellar mass of
galaxies.  At the median redshift of the SWIRE survey, this
corresponds to the observed 3.6~\micron\ band, thus a sample of
galaxies selected at 3.6~\micron\ is essentially selected on the basis
of the total mass in stars.

\begin{figure}[t]
\includegraphics[width=0.43\hsize]{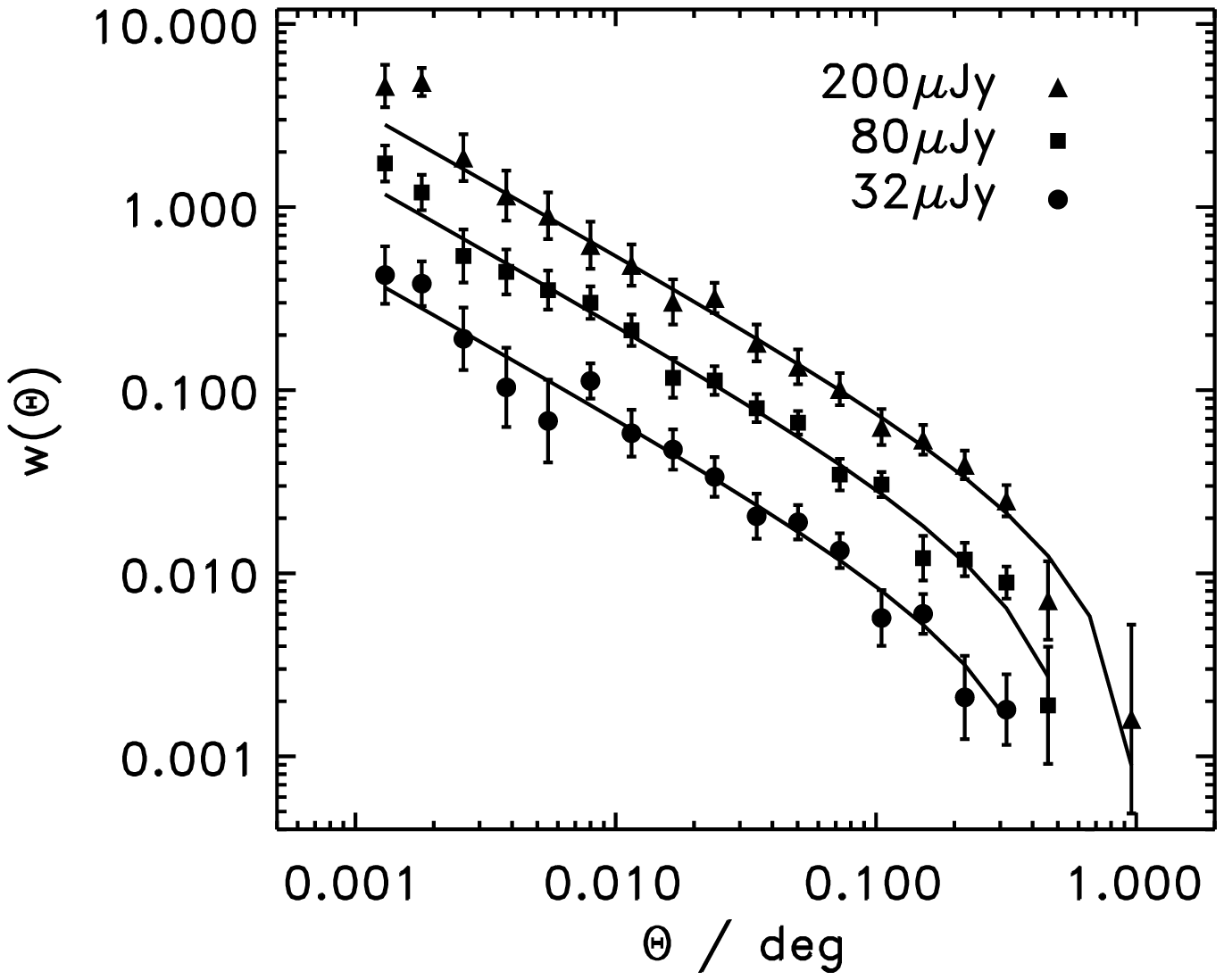}
\includegraphics[width=0.54\hsize]{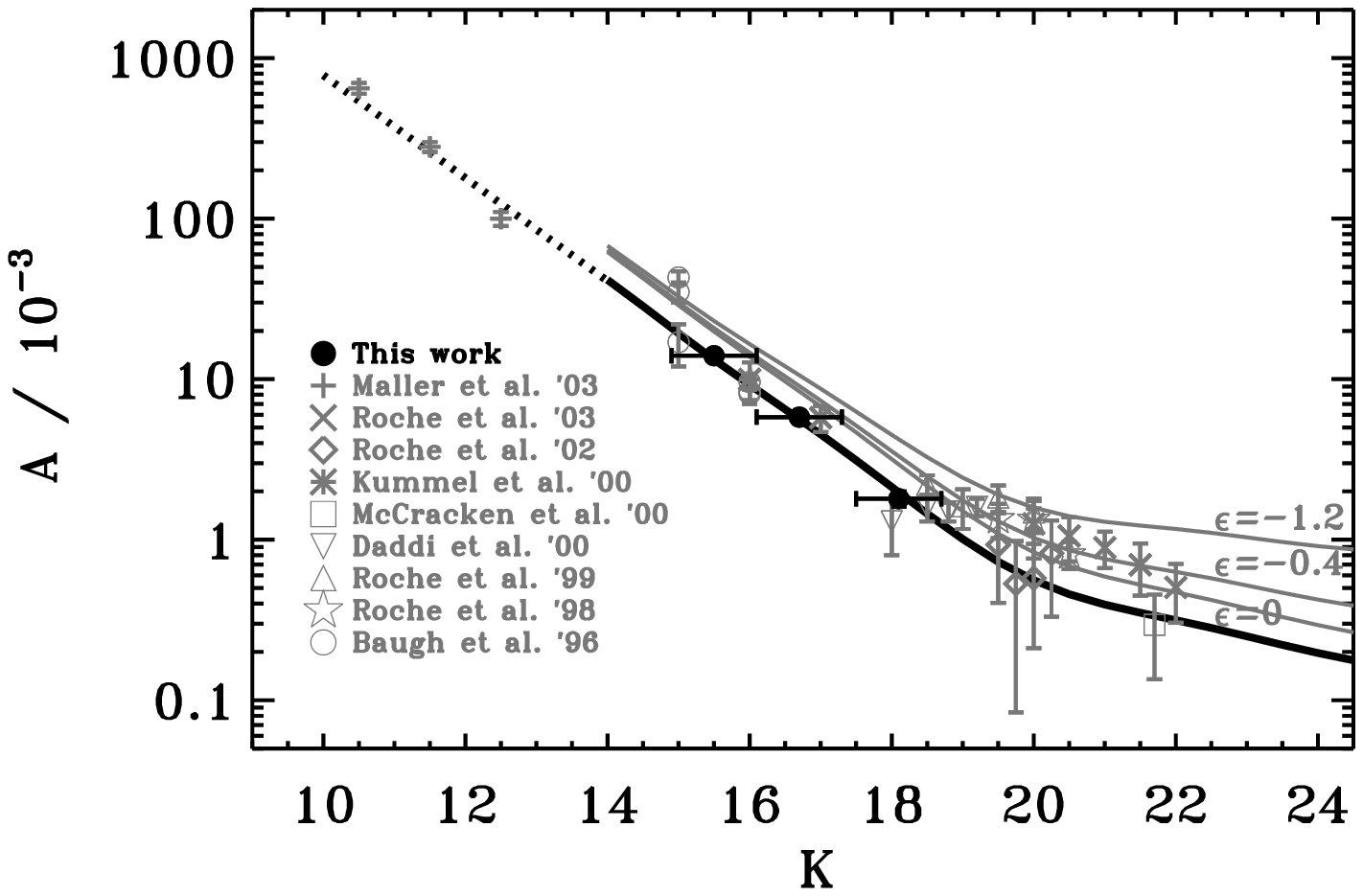}
\caption{(a) Two-point angular correlation functions at
  3.6~\micron\ for three flux limited samples within the SWIRE
  ELAIS-N1 field with $S_{3.6}\ge 32$~$\mu$Jy, 80~$\mu$Jy and
  200~$\mu$Jy.  The best-fitting power-law models,
  $w(\theta)=A\theta^{-0.8}-C$, are shown for each sample.  (b) The
  amplitude, $A$, of the angular correlation function as a function of
  $K$-band magnitude, compared with the evolving models (light lines)
  of \citet{Roche03}.  The heavy/dotted line is the $\epsilon=0$ model
  normalized to the SWIRE data (filled circles).}
\end{figure}

The ELAIS (European Large-Area ISO [Infrared Space Observatory]
Survey) N1 field has an area of $\sim$9~sq.~deg.\ and has been
observed in all seven Spitzer bands.  The Version 1.0 source catalogue
\citep{Surace04} has been used to select three flux-limited samples at
3.6~\micron.  A faint sample was selected such that the flux limit of
$S_{3.6}\ge 32$~$\mu$Jy was well-above the survey detection limit,
ensuring a high source reliability (signal-to-noise $> 40$) and a
uniform selection function; this sample covers an area of 1~sq.~deg.
A medium sample with $S_{3.6}\ge 80$~$\mu$Jy covers 2~sq.~deg.\ and a
bright sample has $S_{3.6}\ge 200$~$\mu$Jy over 8~sq.~deg.  Note that
the faint and medium samples are contained within the bright sample
and are not independent.  Following \citet{Oliver04}, sources
were removed from the initial catalogue if they were identified as
point sources in the Two-Micron All Sky Survey (2MASS) or had the
3.6/4.5-\micron\ colours of stars.  The remaining stellar
contamination was estimated to be $<3$\%.

The two-point angular correlation function, $w(\theta)$, was
calculated for each of the three samples following the same techniques
as \citet{Oliver04}, using the Landay--Szalay estimator.  The
correlation functions are plotted in figure~1a.  A power-law model,
$w(\theta)=A\theta^{1-\gamma}-C$, was fitted to the data over
$\theta>0.003$~deg (11\arcsec) -- the 6\arcsec\ photometry aperture
limits the reliability of the data on smaller scales.  With a fixed
slope of $\gamma=1.8$, the best-fitting correlation amplitudes, $A$,
were $(1.8\pm0.2)\times10^{-3}$ at 32~$\mu$Jy,
$(5.8\pm0.4)\times10^{-3}$ at 80~$\mu$Jy and
$(13.9\pm1.0)\times10^{-3}$ at 200~$\mu$Jy.

\section{Comparison with {\sc galics} mock catalogues}

{\sc galics} (Galaxies In Cosmological Simulations) is a hybrid model
of galaxy evolution which combines high-resolution N-body simulations
of the dark matter content of the universe with semi-analytic
prescriptions to describe the fate of the baryons within the dark
matter halos \citep{hatton03}.  The simulations have 256$^3$ particles
in a 140~h$_{70}^{-1}$~Mpc box, with a minimum halo mass of
$2\times10^{11}$~M$_\odot$.  Within each halo, some fraction of the
gas mass is cooled and turned into stars which then evolve.  The
spectral energy distributions of these model galaxies are computed by
summing the contribution of all the stars they contain, tracking their
age and metallicity.  A mock catalogue is generated by projecting a
cone through the simulation at a series of timesteps (redshifts), and
calculating the properties of the galaxies `observed' in the cone.
The {\sc galics} project have made available several of these cones,
from which we have constructed mock catalogues of the SWIRE survey, in
1-sq.-deg.\ patches.

\begin{figure}[t]
\includegraphics[width=\hsize]{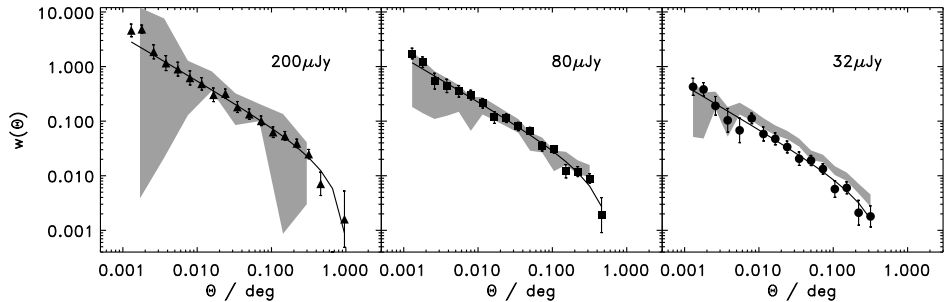}
\caption{The SWIRE 3.6-\micron\ angular correlation functions compared
  with the {\sc galics} simulations for the three flux-limited
  samples.  The shaded regions are the 1-$\sigma$ error bounds on
  $w(\theta)$ from the mock catalogues; the data points and
  best-fitting curves correspond to figure~1a.}
\end{figure}

We have calculated the two-point angular correlation functions for the
{\sc galics} catalogues, using three flux-limited samples
corresponding to the SWIRE data described in the previous section.  In
figure~2 we compare the {\sc galics} correlation functions (shaded
regions) with the SWIRE results (points; figure~1a).  At the brighter
flux limits ($S_{3.6}\ge 80$~$\mu$Jy \& 200~$\mu$Jy) the clustering in
the simulations is consistent with the data.  For a fixed
$\gamma=1.8$, the amplitudes of the best-fitting power-laws differ by
$<2$-$\sigma$ between the data and the simulations in each sample.
The larger uncertainties in the model correlation functions are due to
the small size (1~sq.~deg.) of the simulations compared with the data
(8~sq.~deg.\ at 200~$\mu$Jy) -- there are significantly fewer sources
contributing to the estimation of $w(\theta)$.  In the faintest
sample, $S_{3.6}\ge 32$~$\mu$Jy, the model predicts stronger
clustering than we observe on scales $>1$~arcmin.  The amplitude of
the {\sc galics} correlation function, $A=(2.7\pm0.2)\times10^{-3}$,
differs by $>3$-$\sigma$ from that of the SWIRE data.

\section{Clustering evolution}

The angular correlation function is the projection along the line of
sight of the spatial correlation function, $\xi$, and is dependent on
both the redshift distribution and luminosity function of galaxies in
the survey and on the evolution of the spatial clustering.  At fainter
flux limits the survey probes to higher redshifts and larger volumes,
and this reduces the strength of the projected clustering.  This is
shown in figure~1b where we plot the amplitude, $A$, of the angular
correlation function against the limiting magnitude for a range of
$K$-band surveys \citep{Roche03}.  We used the {\sc galics}
simulations to estimate the equivalent $K$-band limit for our
3.6-\micron-selected samples, and plot these SWIRE clustering
amplitudes on the same figure.  Our new data have much smaller errors
in $A$ and are in agreement with the $K$-band surveys.

Also plotted in figure~1b (light lines) are three models of clustering
evolution, $\xi(r,z)=(r/r_0)^{-\gamma}(1+z)^{-(3+\epsilon)}$ with
$\epsilon=0$, $-0.4$ \& $-1.2$, from \citet{Roche03}.  We note that
these models appear to be offset from the data for $K\le18$, and we
have renormalized the $\epsilon=0$ model to agree with the SWIRE
amplitudes (heavy line).  This renormalized model is consistent with
most of the existing $K<18$ data and, significantly, a linear
interpolation of the model passes through the 2MASS data points at
$K<14$ (dotted line).  

We make several conclusions based on these results.  First, the
correlation length $r_0$ (or the amplitude $A$) of stellar mass
selected samples (at $K$-band \& 3.6~\micron) appears to be smaller
than that of the $I$-band selected sample used in the models
\citep{Roche03}.  Second, the scatter in the ground-based $K$-band
data limits their ability to discriminate between the models
(figure~1b).  The SWIRE flux limit, $S_{3.6}=3.7$~$\mu$Jy, corresponds
to an equivalent $K\simeq20.5$ and we will soon measure the
correlation function to this depth, with smaller errors than the
existing data.  Third, the observed amplitude of the correlation
function at faint flux limits is inconsistent with the {\sc galics}
catalogues (figure~2) and appears to be more complex than the
power-law models of \citet{Roche03} in figure~1b.  Detailed modelling
of the redshift distribution, luminosity function and clustering
evolution is being developed, which will enable us to gain further
understanding of the observed galaxy clustering.

\acknowledgements Thanks to: Carol Lonsdale and the SWIRE team for all
their work on the survey; Eduardo Gonzales-Solares for software to
estimate $w(\theta)$; Nathan Roche for providing us with his models;
and the {\sc galics} project for making their mock catalogues public
(see {\tt http://galics.iap.fr/}).


\end{document}